\documentclass{article}

\usepackage{PRIMEarxiv}

\usepackage[utf8]{inputenc} % allow utf-8 input
\usepackage[T1]{fontenc}    % use 8-bit T1 fonts
\usepackage{hyperref}       % hyperlinks
\usepackage{url}            % simple URL typesetting
\usepackage{booktabs}       % professional-quality tables
\usepackage{amsfonts}       % blackboard math symbols
\usepackage{amsmath}
\usepackage{nicefrac}       % compact symbols for 1/2, etc.
\usepackage{microtype}      % microtypography
\usepackage{fancyhdr}       % header
\usepackage{graphicx}       % graphics
\usepackage{pifont}
\graphicspath{{media/}}     % organize your images and other figures under media/ folder

\title{Data-driven approach for diagnostic analysis of dynamic bottlenecks in serial manufacturing systems}

\author{
  Nikolai West \\
   \\
  Institute of Production Systems \\
  Technical University Dortmund \\
  Dortmund, Germany\\
   \\
   RIF Institute for Research and Transfer e.V.\\
   Work and Production Systems\\
   Dortmund, Germany\\
   \\
  \texttt{nikolai.west@tu-dortmund.de}\\
  0000-0002-3657-0211\\
  \And
  Jörn Schwenken\\
   \\
  Institute of Production Systems \\
  Technical University Dortmund \\
  Dortmund, Germany\\
   \\
   RIF Institute for Research and Transfer e.V.\\
   Work and Production Systems\\
   Dortmund, Germany\\
   \\
  \texttt{joern.schwenken@ips.tu-dortmund.de}\\
  0000-0003-2549-9664\\
  \AND
  Jochen Deuse \\
  \\
  Institute of Production Systems \\
  Technical University Dortmund \\
  Dortmund, Germany\\
  \\
  University of Technology Sydney\\
  Advanced Manufacturing\\
  Sydney, Australia\\
  \\
  \texttt{jochen.deuse@ips.tu-dortmund.de}\\
  0000-0003-4066-4357\\
}

\begin{document}
\maketitle

\begin{abstract}
A variety of established approaches exist for the detection of dynamic bottlenecks. Furthermore, the prediction of bottlenecks is experiencing a growing scientific interest, quantifiable by the increasing number of publications in recent years. Neglected, on the other hand, is the diagnosis of occurring bottlenecks. Detection methods may determine the current location of a bottleneck, while predictive approaches may indicate the location of an upcoming bottleneck. However, mere knowledge of current and future bottlenecks does not enable concrete actions to be taken to avoid the bottlenecks, nor does it open up any immediate advantage for manufacturing companies. Since small and medium-sized companies in particular have limited resources, they cannot implement improvement measures for every bottleneck that occurs. Due to the shifts of dynamic bottlenecks, the selection of the most suitable stations in the value stream becomes more difficult. This paper therefore contributes to the neglected field of bottleneck diagnosis. First, we propose two data-driven metrics, relative bottleneck frequency and relative bottleneck severity, which allow a quantitative assessment of the respective bottleneck situations. For validation purposes, we apply these metrics in nine selected scenarios generated using discrete event simulation in a value stream with a serial manufacturing line. Finally, we evaluate and discuss the results. 
\end{abstract}


\section{Introduction}
\label{sec:introduction}
The diagnosis of throughput-limiting bottlenecks is essential for manufacturing companies that want to maintain a high degree of production efficiency. According to the \textbf{Theory of Constraints (TOC)}, every system is inevitable limited by a bottleneck, which must be identified and optimized to improve the systems overall output \cite{Goldratt1984}. Since the TOC is considered universally applicable, its rules apply to manufacturing systems as well. TOC has been the subject of on-going research efforts for several decades. In all but the most simple manufacturing systems, bottlenecks are not static, but change dynamically \cite{Roser2002}. A particular challenge arises due to this shifting behavior of manufacturing bottlenecks. Due to the increasing demand for flexibility and due to the rising complexity in interconnected value streams, the variability in modern value streams increases as well. Shifting behavior is considered an important underlying principle for dealing with manufacturing bottlenecks. Thus, most scientific work has primarily dealt with two questions:

\begin{itemize}
    \item \textbf{Detection:} Where is the bottleneck at this moment?
    \item \textbf{Prediction:} Where is the bottleneck going to be next?
\end{itemize}

Several methods exist to detect the current location of a bottleneck \cite{Betterton2012}. These methods are either based on a momentary snapshot of the current conditions of the manufacturing system or they are based on an averaged evaluation of a given period of past system behavior \cite{West2022a}. Due to its good applicability for dynamic bottlenecks \cite{West2022b}, we are going to apply the \textbf{Active Period Method (APM)} in this paper \cite{Roser2002}. APM considers the station with the longest active operating time as the current bottleneck. We further elaborate the usage of the method in \textbf{Section \ref{sec:apm}}.

Due to the emerging capabilities in analyzing large volumes of data using machine learning and artificial intelligence, research in recent years mainly aims to predict the future location of a bottleneck. The evolving possibilities of intelligent and data-driven analyses are being used for this purpose. Even though the research field is still rather young, there are several promising approaches for making data-driven predictions of future bottleneck events \cite{Lin2022, Rocha2022, Bui2021}. Similar to detection, such bottleneck prediction requires data-driven diagnostic tools to evaluate different scenarios. As such, we emphasize the need for bottleneck diagnosis, which has been neglected so far.

\begin{itemize}
    \item \textbf{Diagnosis:} What are the effects of occurring bottlenecks?
\end{itemize}

The remainder of this paper is organized as follows. First, we introduce fundamentals of Bottleneck Analysis (\textbf{Section \ref{sec:fundamentals}}). We discuss a holistic model, bottleneck detection with AMP and the universal need for bottleneck diagnosis. Then, we propose two statistical metrics for bottleneck diagnosis and explain the calculations for bottleneck frequency and severity (\textbf{Section \ref{sec:metrics}}). To validate these metrics, we apply them in a number of simulations. For this purpose, we outline the nine selected scenarios in the next chapter (\textbf{Section \ref{sec:simulation}}). At last, we present the results of the application (\textbf{Section \ref{sec:results}}) and discuss the results with a brief outlook on follow-up work (\textbf{Section \ref{sec:conclusion}}).

\section{Fundamentals}
\label{sec:fundamentals}

\subsection{Four steps of a holistic Bottleneck Analysis}
\label{sec:holistic}

\begin{figure}
  \centering
  \includegraphics[scale=0.15]{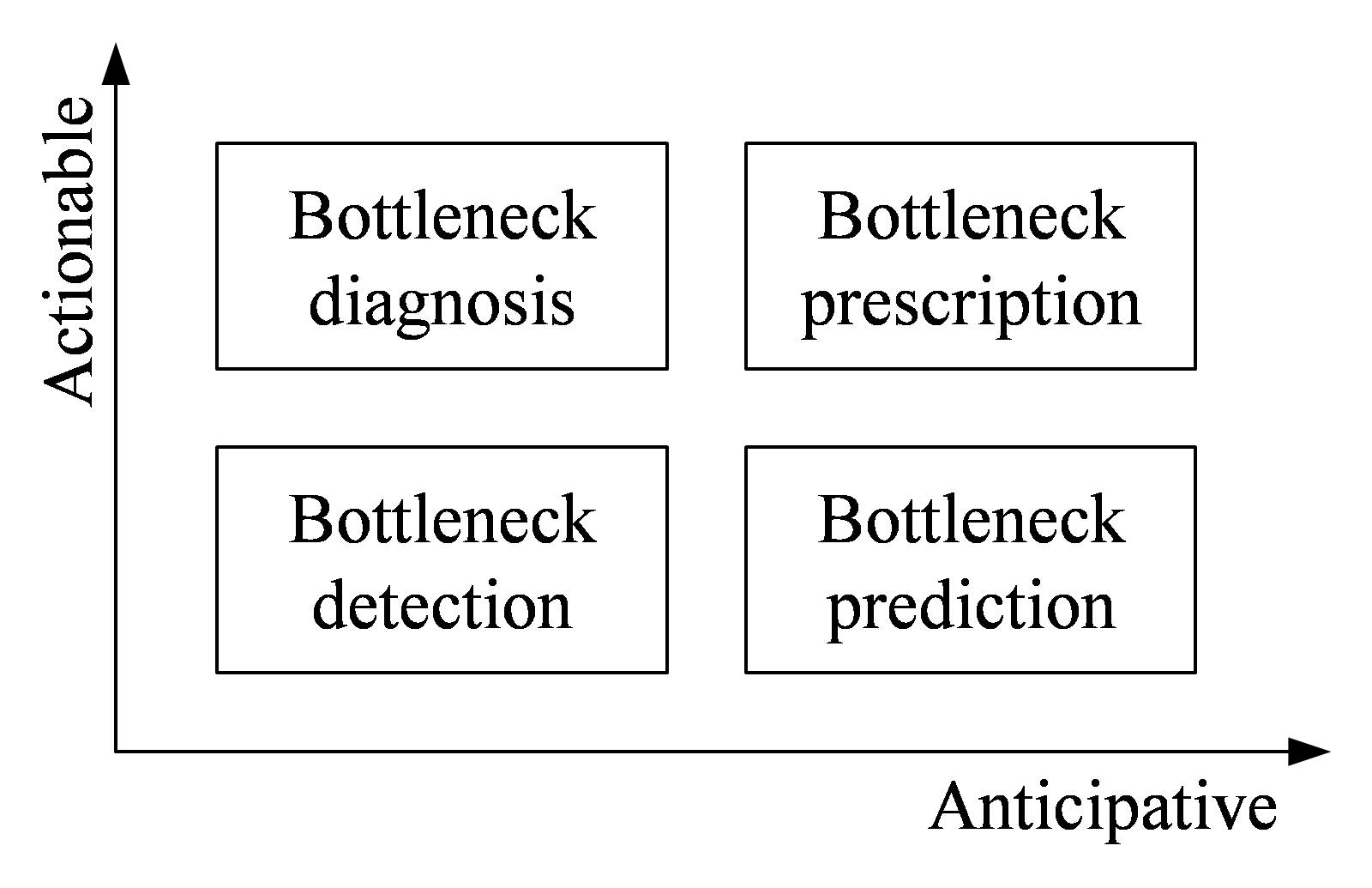}
  \caption{Depiction of the four steps for Bottleneck Analysis in the context of actionable and anticipative results \cite{West2022a}}
  \label{fig:fig1}
\end{figure}

Bottleneck Analysis is a complex process that requires a well-structured methodology to ensure that all necessary tasks are completed in a successive and goal-oriented manner. Within this paper, we follow a holistic model for Bottleneck Analysis to distinguish four major steps: Bottleneck detection, diagnosis, prediction, and prescription \cite{West2022a}. \textbf{Figure \ref{fig:fig1}} depicts the four steps in the context of actionable and anticipative results. 

The first step, \textbf{Bottleneck Detection}, is akin to descriptive analytics and fulfills a declarative task during Bottleneck Analysis. It involves the concise determination of the current bottleneck in the manufacturing system, based on information about the system's status. To identify bottlenecks, different methods may be applied, such as machine states, buffer level, or process times \cite{Betterton2012, Yu2016}. We propose using a Momentary Value Method to identify bottlenecks in real-time, as they can identify a bottleneck at any point in time, whereas Average Value Methods rely on arbitrary periods of system observations \cite{West2022b}.

The second step, \textbf{Bottleneck Diagnosis}, originates from diagnostic analysis and primarily addresses an assessment task during Bottleneck Analysis. The goal of this step is to evaluate and assess the cause and effect of observed bottlenecks \cite{West2022a}. The diagnosis of bottlenecks is the focus of this paper and will be explained in more detail later on. Our main goal is to standardize future work by developing two simple metrics to quantify bottleneck impacts. In the past, diagnosis was usually either neglected or performed based on qualitative estimates that included consultation with experts of the manufacturing system.

The third step, \textbf{Bottleneck Prediction}, relates to predictive analytics and fulfills an anticipatory task during Bottleneck Analysis. Its goal is to determine the future performance within complex manufacturing systems. The prediction step requires prior implementation of a real-time bottleneck detection system and can be achieved by predicting future machine states, buffer level, and numeric forecast of the development of interdeparture time variances. However, further research is necessary to develop prediction methods for Bottleneck Analysis. 

Lastly, the fourth step, \textbf{Bottleneck Prescription}, is similar to prescriptive analytics and fulfills a preemptive task during Bottleneck Analysis. It involves using intelligent system control to mitigate bottlenecks, such as bottleneck-centered production control systems. Although most approaches consider systems that are geared to past system states, we suggest using Bottleneck Diagnosis and Prediction to ensure that the system is up-to-date and responsive to changing conditions.

In conclusion, the four steps help to manage and utilize the potential of Bottleneck Analysis in manufacturing. The holistic model may still be novel, but it has been included in a number of works in the past years \cite{Mahmoodi2022}. For a more extensive review of the bottleneck literature, we refer to the excellent literature review in \cite{Subramaniyan2021}. For this paper, we conclude that Bottleneck Analysis is a continuous process that requires constant attention to ensure the efficient operation of a manufacturing system.

\subsection{Bottleneck detection using the longest active period}
\label{sec:apm}

The method presented below is one possible approach to detect bottlenecks. As mentioned in the introduction, there are several different methods to detect bottlenecks. For a detailed comparison of the methods, we therefore refer to \cite{West2022b} and \cite{Zhao2014}.

\begin{figure}
  \centering
  \includegraphics[scale=1.0]{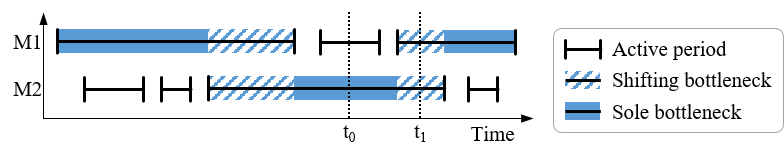}
  \caption{Exemplary visualization of the active periods for two stations $M1$ and $M2$ with a shifting bottleneck \cite{Roser2002}}
  \label{fig:fig2}
\end{figure}

The Active Period Method (APM) is a method used for bottleneck detection in manufacturing systems. According to the APM, a bottleneck is the station in the value stream that has been working the longest without interruptions. This duration is then called the active period of the station. A station is considered active if it is processing products as defined by the production program. On the other hand, a station is considered to be inactive if it is waiting due to buffer-related starvation or blockage \cite{Roser2002, Roser2002b}. A station is blocked if the downstream buffer is filled to the maximum. Likewise, a station is starved if the upstream buffer is empty and cannot supply another part or product to the next station. APM also incorporates shifting states of a bottleneck to determine whether a station is the sole bottleneck or a shifting bottleneck. Shifting behavior occurs at the overlap of the current and the subsequent bottleneck. Figure \ref{fig:fig2} shows an illustrative example of the active periods for two stations $M1$ and $M2$. Here, at $t_0$ the station $M2$ is the sole bottleneck. While at $t_1$ both $M1$ and $M2$ have become shifting bottlenecks \cite{Roser2002}. 

Summarizing, the main advantage of APM is that it is easy to apply and requires neither extensive analysis nor complex mathematical modeling. In addition, as a non-invasive method, APM does not require shutting down the manufacturing system to collect data. Overall, APM is a simple and useful method to detect bottlenecks in manufacturing systems \cite{West2022b}. 

\subsection{On the need for metrics for bottleneck diagnosis}
\label{sec:diagnosis}

While both bottleneck detection and prediction receive considerable attention in the scientific discourse, bottleneck diagnosis has been widely neglected. Nevertheless, some approaches to performing a diagnosis do exist. In an early work, \cite{Zhou2006} propagate a simple visual evaluation for the Bottleneck Analysis to determine the effects of the bottleneck stations. Using an average-based performance metric, such as the utilization, it was then possible to quantify the bottleneck effect of individual stations. In a promising paper, \cite{Subramaniyan2020} use a clustering-based approach to prioritize maintenance activities. Through the unsupervised approach, maintenance practitioners are provided with information on the maintenance-related diagnostic insights into bottlenecks. \cite{Kumbar2023} include a diagnostic examination in their digital twin based framework for throughput bottlenecks. This is intended to prepare manufacturing companies for the digital transformation and to enable them to ‘utilize the wealth of enterprise information’. Common to all previous approaches is a lack of a metric to easily compare bottleneck effects. Two such metrics are developed in this paper and further elaborated in \textbf{Section \ref{sec:metrics}}.

As shown in \textbf{Figure \ref{fig:fig1}}, both bottleneck detection and prediction have a limited degree of actionability. In contrast, a targeted diagnosis of existing bottlenecks opens up the possibility of prioritizing measures to reduce the overall effects of bottlenecks. This is especially relevant in scenarios characterized by dynamic bottlenecks and a finite amount of resources:

\begin{itemize}
    \item \textbf{Dynamic behavior:} Since any station can become the bottleneck in systems with dynamic shifting, the focus of improvement activities must be changed on a regular basis.
    \item \textbf{Finite resources:} Since the resources that are available for improvement are limited, the activities must be concentrated on the stations most heavily affected by bottleneck behavior.
\end{itemize}

Summarizing, a successful bottleneck diagnosis is a critical step in improving the performance of a system or process, and it is essential for organizations that want to remain competitive and efficient in today's fast-paced business environment.

\section{Metric proposal}
\label{sec:metrics}

As previously mentioned, the focus of this paper is on two simple metrics for evaluating bottleneck behavior. The metrics are intended to enable users to evaluate detected bottlenecks in order to select improvement measures in a targeted manner and maximize the impact of these measures. Therefore, both metrics are used in deterministic calculations that we explain next.

\subsection{Relative bottleneck frequency}
We refer to the first metric used to diagnose bottlenecks as \textbf{relative bottleneck frequency} or $rbf$. This metric uses the intuitive approach of determining how often an individual station or machine occurs as a bottleneck during the period under consideration. This simple idea builds on the fundamental assumption of TOC that there can only be one bottleneck at any given time. Let $S$ be a station in a manufacturing system. Then $rbf_S$ represents the relative bottleneck frequency for $S$. The analysis takes place for a defined period of time t of a fixed length n. At each point in time $t_i$, the current bottleneck of the entire system needs to be determined. Conceptually, there are no restrictions in the choice of methods as long as they clearly identify a single station as a bottleneck for each time point. Then, we can determine how often S appears as a bottleneck during t. We call this value the (total) bottleneck frequency $bf_S$. It is calculated as the quotient of the number of times at which S is the bottleneck during t and n, the total length of t. \textbf{Equation \ref{eq:eq1}} and \textbf{Equation \ref{eq:eq2}} show the mathematical calculation of the relative bottleneck frequency. 

\begin{equation}
\label{eq:eq1}
\mathbf{S} = \sum_{\substack{0 < i < n}} \begin{cases}
1 & : \ \text{if } S \text{ is bottleneck at } t_i,\\
0 & : \ \text{else}.
\end{cases}
\end{equation}

\begin{equation}
\label{eq:eq2}
rbf_S = \frac{bf_s}{n}
\end{equation}

The following applies to the value range $rbf_S$ in ${(0,...,1)}$. During bottleneck diagnosis, $rbf$ has to be determined for each S. The higher $rbf_S$, the more frequently $S$ occurs as bottleneck. The case of $rbf$ having a value of 1 implies a static bottleneck on S, while an $rbf_S$ of 0 corresponds to S never becoming a bottleneck. In addition, the sum of the $rbf_S$ of all stations in the value stream under consideration must always add up to 1. 

\subsection{Relative bottleneck severity}

While relative bottleneck severity is well suited to evaluate an observation period, it fails when examining individual points in time. This dilemma is already known from the detection of bottlenecks. Here, momentary value methods in particular prove useful because, in contrast to average value methods, they do not require any information about past system states. For this reason, we propose a second metric, the \textbf{relative bottleneck severity} or $rbs$, that is based only on the current system state. 

Like $rbf$, $rbs$ is determined for each station $S$ in the value stream. The determination of the severity is based on the respective characteristic of the selected bottleneck detection method. For the purpose of illustration, we will explain the calculation of $rbs$ using the APM as an example. To apply the APM, the duration of the active operating period of all stations is known at each point in time. We refer to this duration as $bs_s$ for each $S$. Since the bottleneck station is characterized by the longest active operating period, we refer to this value as $bs_{\mathrm{BN}}$.

\begin{equation}
rbs_S = \frac{bs_S}{bs_{BN}}
\label{eq:eq3}
\end{equation}

$rbs$ has the same range of values as $rbf$ of ${(0,...,1)}$, but the sum of $rbs_S$ for all stations $S$ can be greater than 1 (\textbf{Equation \ref{eq:eq3}}). The station occurring as the bottleneck at the time $t_i$ of the analysis has an $rbs$ value of 1, because $bs_S$$=$$bs_\mathrm{BN}$ must always be valid for this station. The $rbs$ value of the other stations then indicates the severity of their current impact on the bottleneck. The closer the $rbs$ value is to 1, the more severe is the impact on the system.

While we have explained the calculation of $rbs$ for APM only, in principle other detection methods can also be used to determine $rbs$. For example, the interdeparture time variance that relies on the stations’ current processing times \cite{Betterton2012} or an adaption of the bottleneck walk that considers the current buffer level before and after each station could be utilized as well. 

\section{Simulation scenarios}
\label{sec:simulation}

To demonstrate the usability of the two metrics $rbf$ and $rbs$, we apply them in nine exemplary scenarios in this section. We generated the manufacturing data ourselves using a discrete event simulation. In order to obtain comprehensible results from the metrics, we apply them in simple flow lines with seven fully interlinked stations. We separate nine scenarios in three main categories ($S1$, $S2$, $S3$), each consisting of three scenarios.

\begin{itemize}
    \item \textbf{S1}: No station receives an increased process time pt, instead the variability var of the process times is changed in each scenario by 25\% ($S1\mbox{-}1$, $S1\mbox{-}2$ and $S1\mbox{-}3$).
    \item \textbf{S2}: One station receives an additional increase of 12.5\% to its average process time, and we change the location of this station in each scenario ($S2\mbox{-}1$, $S2\mbox{-}2$ and $S2\mbox{-}3$).
    \item \textbf{S3}: Two stations receives an additional increase of 12.5\% to their average process times, and we change the location of these stations in each scenario ($S3\mbox{-}1$, $S3\mbox{-}2$ and $S3\mbox{-}3$). 
\end{itemize}

To be able to represent the scenarios in later representations we use a simplified notation. A '$\square$' corresponds to a simple station, while a '\ding{110}' corresponds to a station with additional process time. Buffers between stations are represented by a '--'.

Table \ref{tab:table1} shows the nine scenarios in our simplified notation. Furthermore, the simulation required several assumptions. We set the process time pt of every station to 2.00. Every modified station \ding{110} in all scenarios of $S2$ and $S3$ then has a $pt$ of $2.25$. Furthermore, the maximum capacity of all buffers is set to 5 units and the system’s boundaries as set to infinite, providing an unlimited supply of parts and demand for products. Each simulation scenario is run 10 times. The simulation receives a settling time of 2,000 time steps, which are then removed from the analysis. Each scenario corresponds to a one-week observation period with 10,080 individual observations each.

\begin{table}
  \centering
  \caption{Tabular representation of the nine scenarios in the simplified value stream notation}
  \label{tab:table1}
  \renewcommand{\arraystretch}{1.2}
  \begin{tabular}{@{} c @{\quad} c @{\quad} c @{}}
    \hline
    \multicolumn{2}{c}{Scenario} \\
    \cline{2-3}
    Name & Setup of the manufacturing line & Variability \\
    \hline
    S1-1 & $\square$\textminus$\square$\textminus$\square$\textminus$\square$\textminus$\square$\textminus$\square$\textminus$\square$ & Low    \\
    S1-2 & $\square$\textminus$\square$\textminus$\square$\textminus$\square$\textminus$\square$\textminus$\square$\textminus$\square$ & Medium \\
    S1-3 & $\square$\textminus$\square$\textminus$\square$\textminus$\square$\textminus$\square$\textminus$\square$\textminus$\square$ & High   \\
    \hline
    S2-1 & $\square$\textminus\ding{110}\textminus$\square$\textminus$\square$\textminus$\square$\textminus$\square$\textminus$\square$ & Medium \\
    S2-2 & $\square$\textminus$\square$\textminus$\square$\textminus\ding{110}\textminus$\square$\textminus$\square$\textminus$\square$ & Medium \\
    S2-3 & $\square$\textminus$\square$\textminus$\square$\textminus$\square$\textminus$\square$\textminus\ding{110}\textminus$\square$ & Medium \\
    \hline
    S3-1 & $\square$\textminus$\square$\textminus\ding{110}\textminus$\square$\textminus\ding{110}\textminus$\square$\textminus$\square$ & Medium \\
    S3-2 & $\square$\textminus\ding{110}\textminus$\square$\textminus$\square$\textminus$\square$\textminus\ding{110}\textminus$\square$ & Medium \\
    S3-3 & \ding{110}\textminus$\square$\textminus$\square$\textminus$\square$\textminus$\square$\textminus$\square$\textminus\ding{110} & Medium \\
    \hline
  \end{tabular}
\end{table}

The simulation scenarios are implemented using \verb+simpy+ (v4.0.1), a library for discrete event simulation in Python \cite{SimPy}. The raw data of the 90 simulation runs (from nine scenarios with ten simulation runs each) on which the analysis is based, as well as the code used to produce the following results, can be found in the publicly available project repository for this paper.

\begin{center}
  \url{https://github.com/nikolaiwest/2023_bottleneck_diagnosis_arxiv}
\end{center}

\begin{figure}
  \centering
  \includegraphics[scale=0.75]{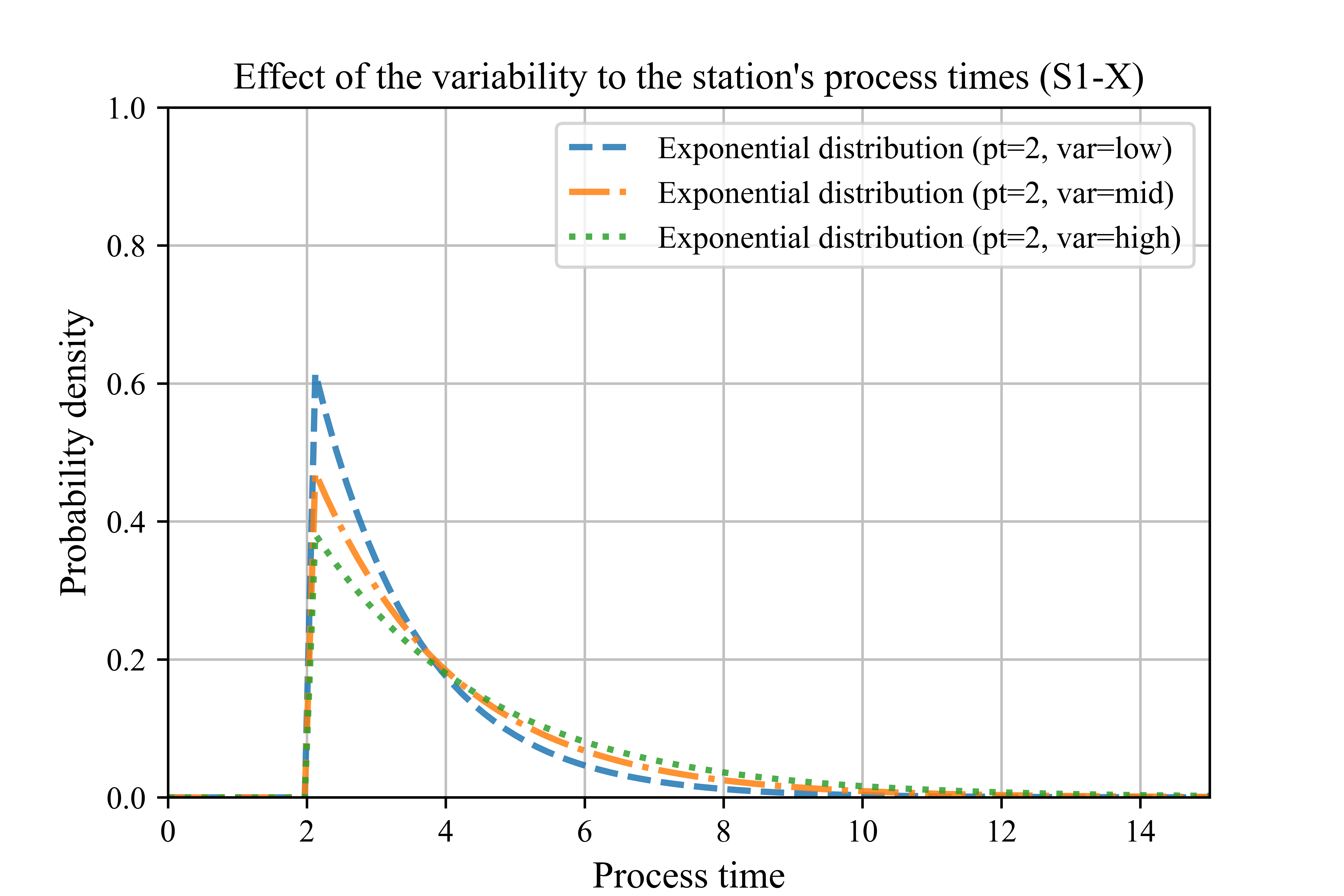}
  \caption{Visualization of the effect of the process time distribution and the applied variability}
  \label{fig:fig3}
\end{figure}

\begin{figure}
  \centering
  \includegraphics[scale=0.75]{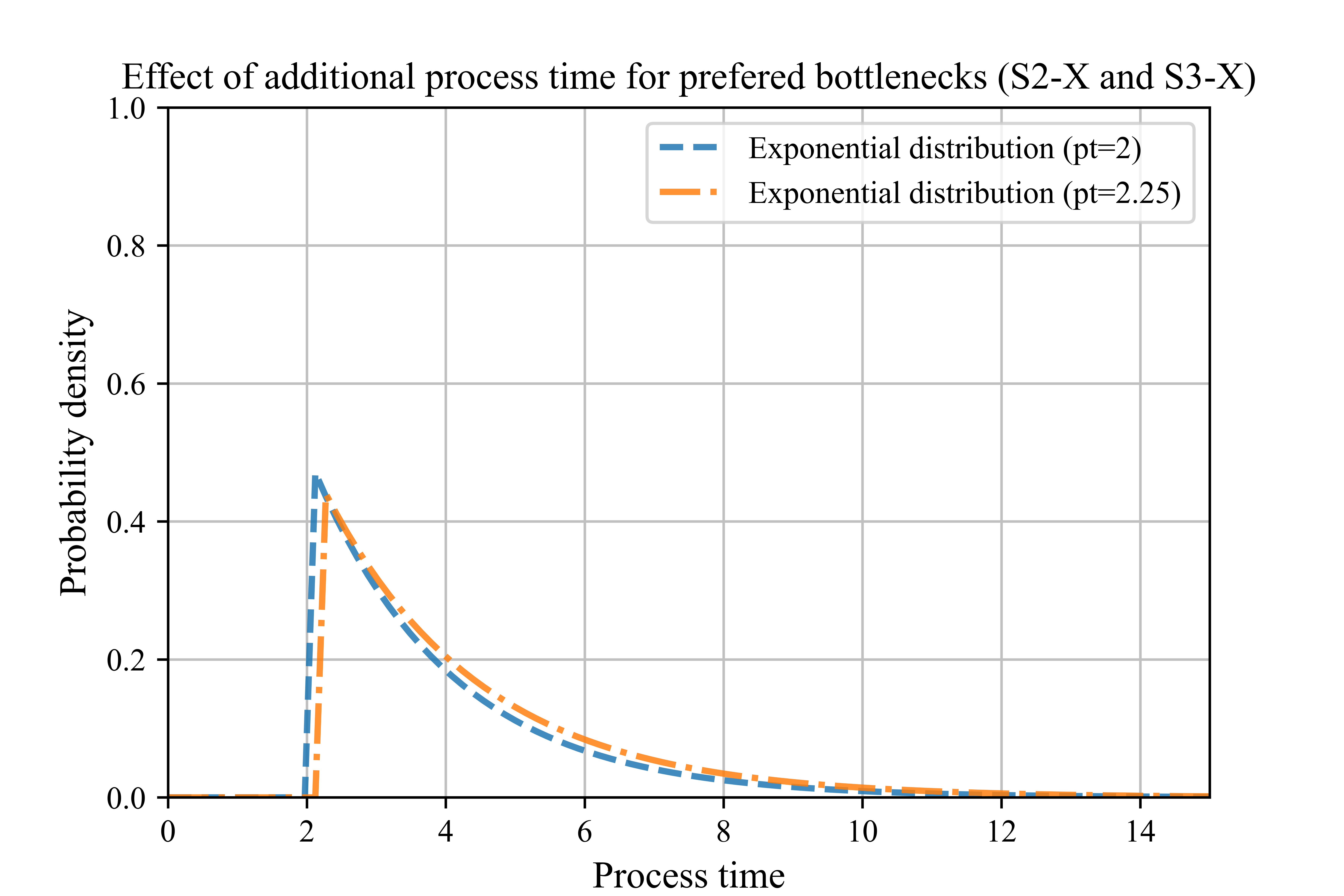}
  \caption{Visualization of the effect of the additional process time for targeted bottleneck stations}
  \label{fig:fig4}
\end{figure}

To achieve a realistic system behavior, variability must be applied to the stations’ process times. For this purpose, we use an exponential distribution, since it most closely corresponds to stations affected by failure or delay. \textbf{Figure \ref{fig:fig3}} and \textbf{Figure \ref{fig:fig4}} serve to illustrate the exponential distribution of the process times. An increase in variability leads to a more frequent occurrence of longer process times. This corresponds to effective downtime due to unforeseen influences. Naturally, the 12.5\% increase in process times leads to a tendency for process times to shift towards longer times. The expected times are always higher than they would be without the process time addition. While we do not anticipate a dominant bottleneck in $S1$, we expect that in $S2$ and $S3$, the \ding{110} stations will increasingly stand out as bottlenecks.

\section{Results}
\label{sec:results}

\begin{figure}
  \centering
  \includegraphics[scale=0.75]{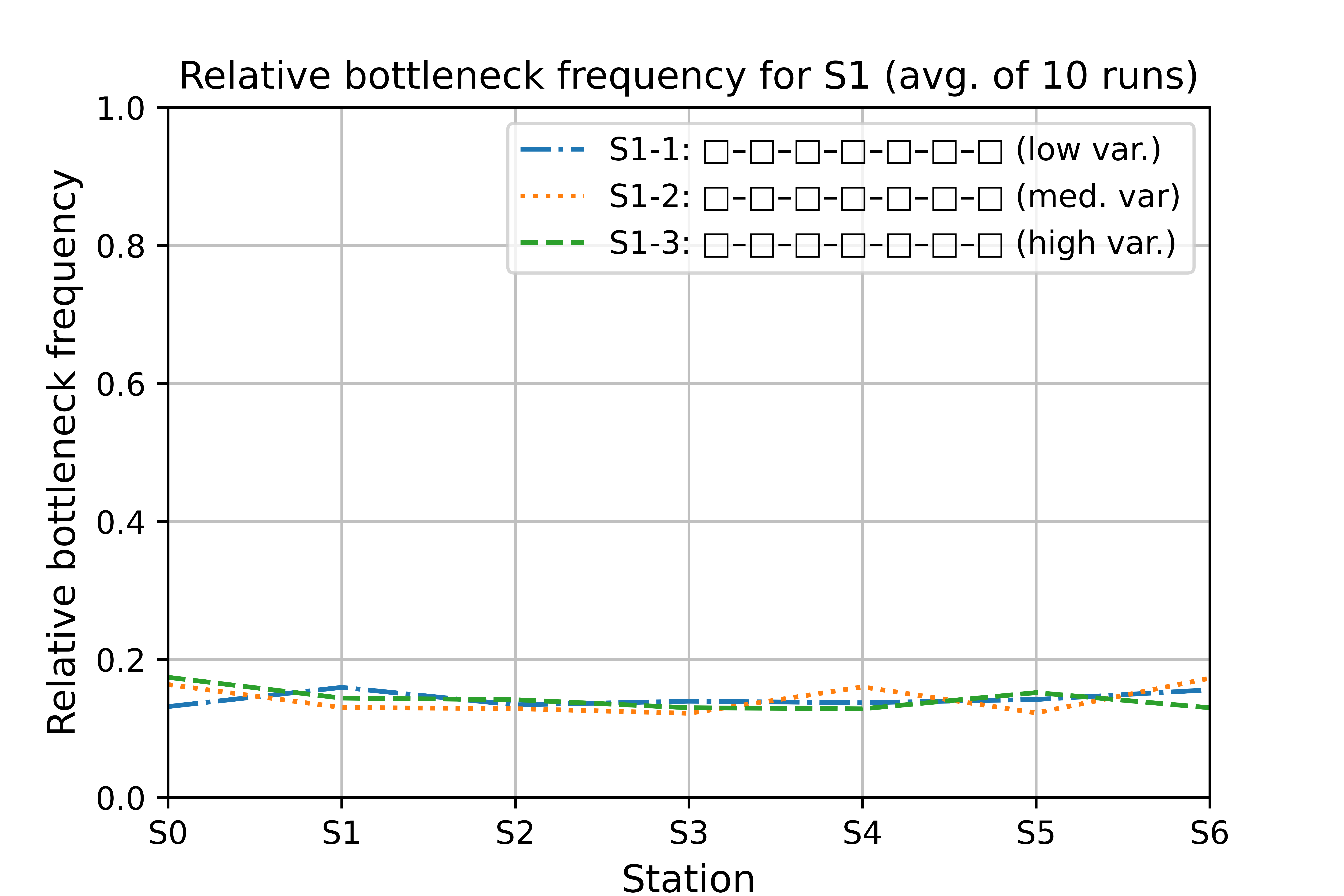}
  \caption{Relative bottleneck frequency for $S1\mbox{-}1$, $S1\mbox{-}2$ and $S1\mbox{-}3$}
  \label{fig:fig5}
\end{figure}

\begin{figure}
  \centering
  \includegraphics[scale=0.75]{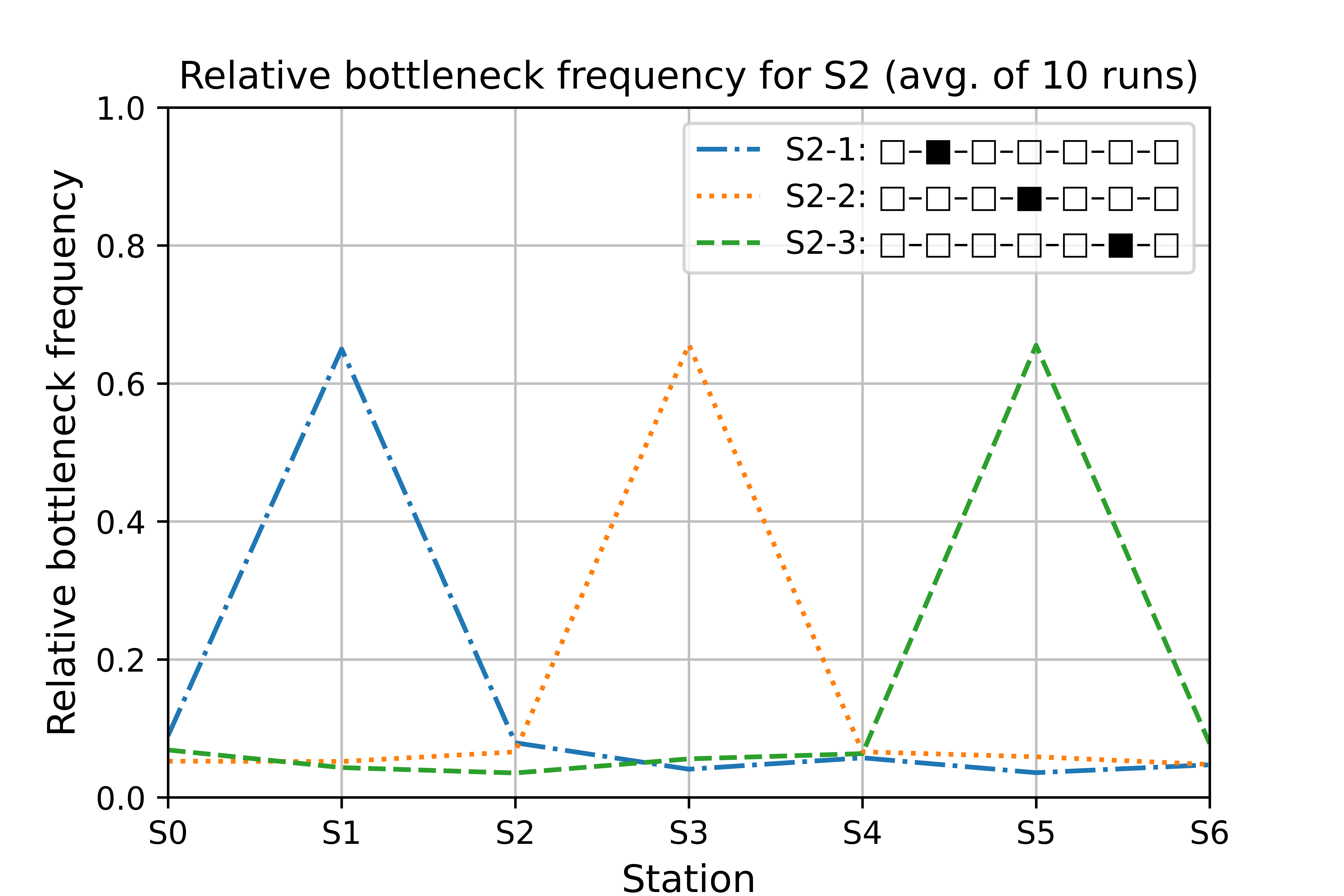}
  \caption{Relative bottleneck frequency for $S2\mbox{-}1$, $S2\mbox{-}2$ and $S2\mbox{-}3$}
  \label{fig:fig6}
\end{figure}

\begin{figure}
  \centering
  \includegraphics[scale=0.75]{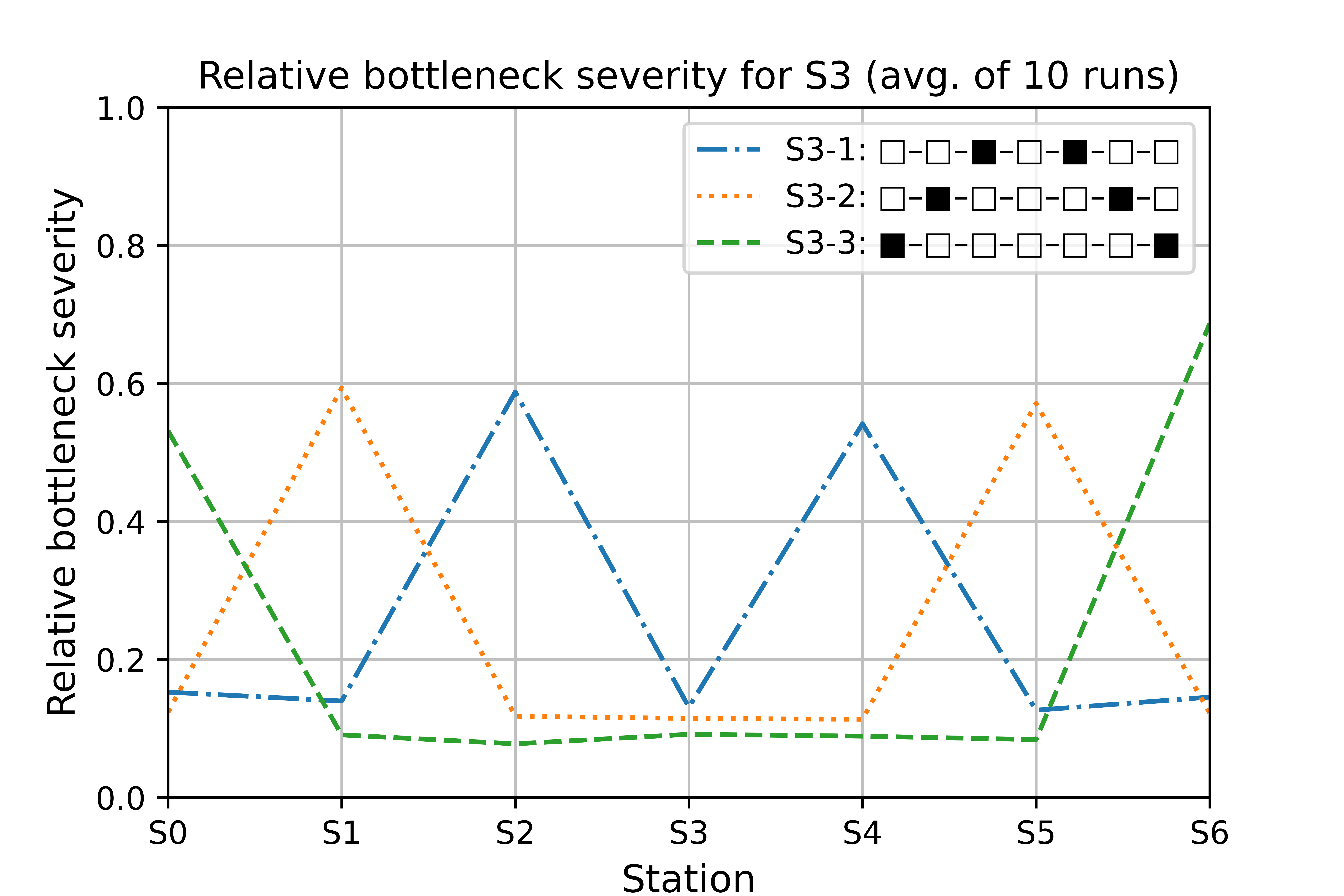}
  \caption{Relative bottleneck frequency for $S3\mbox{-}1$, $S3\mbox{-}2$ and $S3\mbox{-}3$}
  \label{fig:fig7}
\end{figure}

First, we consider the evaluation of the relative bottleneck frequencies. For this purpose, \textbf{Figure \ref{fig:fig5}}, \textbf{Figure \ref{fig:fig6}} and \textbf{Figure \ref{fig:fig7}} show the three scenarios of the main categories, averaged for ten simulation runs. As expected, in no scenario of group $S1$ does a station have a higher tendency to show bottleneck behavior. Despite process times influenced by variance, the $rbf$ for the three scenarios is about $0.15$ for all seven stations which reflects a random spread.

The case is different for the six scenarios of the groups $S2$ and $S3$. Again, we show the average results from ten simulation runs, each with 10,000 steps. Here, the stations, marked by \ding{110}, with additional process time can be frequently identified as bottlenecks on the basis of $rbf$. Consequently, $S1$ is highlighted in $S2\mbox{-}1$, $S3$ in $S2\mbox{-}2$, and $S5$ in $S2\mbox{-}3$. With a value of about $0.65$, the three stations in $S2\mbox{-}1$, $S2\mbox{-}2$ and $S2\mbox{-}3$ show the corresponding station as a bottleneck. Similarly, both modified stations in $S3\mbox{-}1$, $S3\mbox{-}2$, and $S3\mbox{-}1$ are characterized by equally frequent bottlenecks. Since there are two alternating bottlenecks, the $rbf$ value of 0.4 is, as expected, below the value of systems with only one main bottleneck. Overall, we summarize that both the expected hypotheses about the system behavior have been met and the $rbf$ metric contributes to a clear detection and quantitative of the bottleneck stations.

\begin{figure}
  \centering
  \includegraphics[scale=0.75]{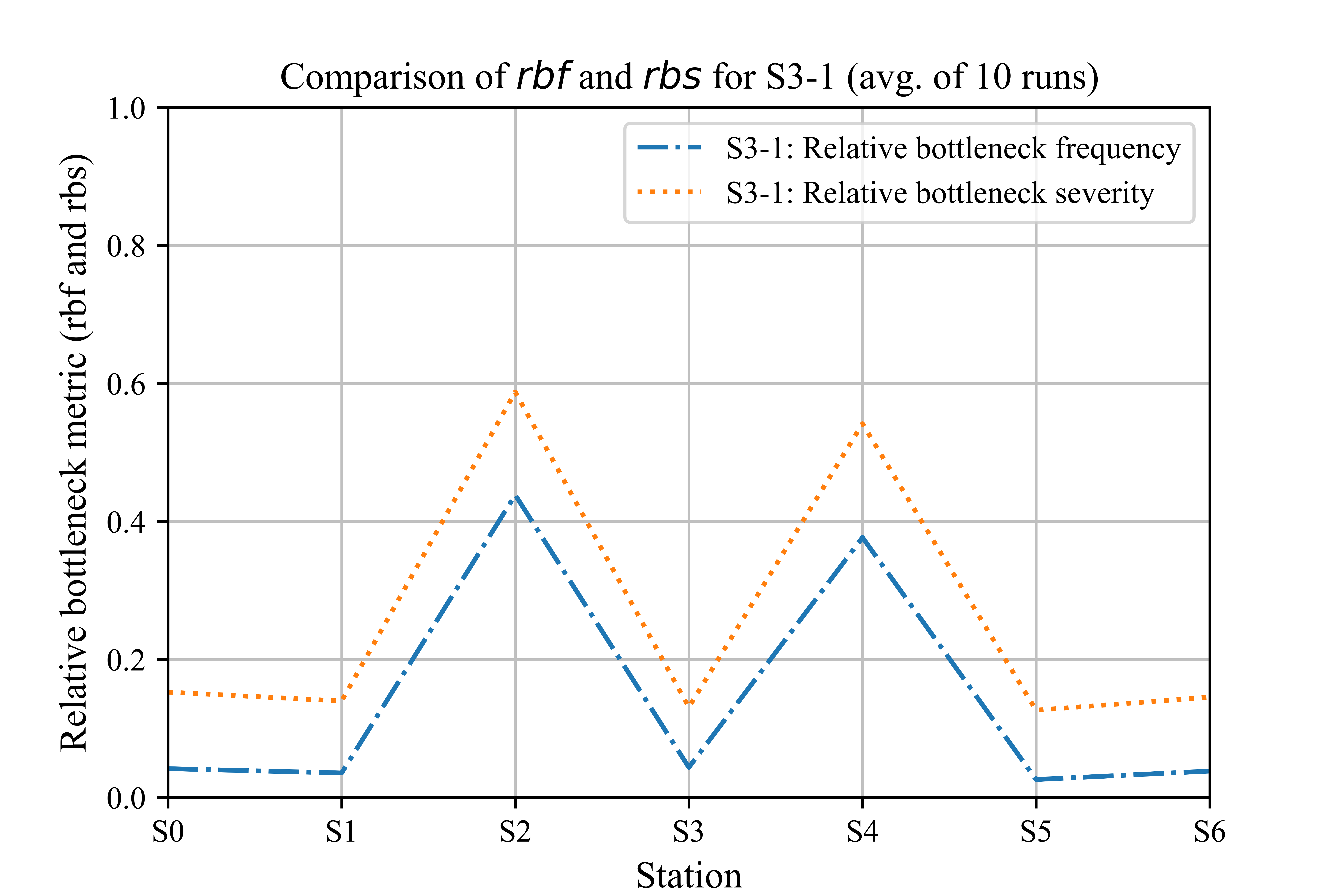}
  \caption{Exemplary comparison of $rbf$ and $rbf$ for $S3\mbox{-}1$}
  \label{fig:fig8}
\end{figure}

We refrain from visualizing all nine scenarios for the corresponding $rbs$ values in this paper for reasons of space. The metric values behave quite similar and can be viewed in the published project repository. Instead, we show a single exemplary comparison for scenario $S3\mbox{-}1$ in \textbf{Figure \ref{fig:fig8}}. The $rbs$ is continuously above the $rbf$, but just so marks the two stations $S2$ and $S4$, changed to $S3\mbox{-}1$, as bottlenecks. In this averaged representation over the entire period, $rbs$ is very similar to $rbf$.

\begin{figure}
  \centering
  \includegraphics[scale=0.75]{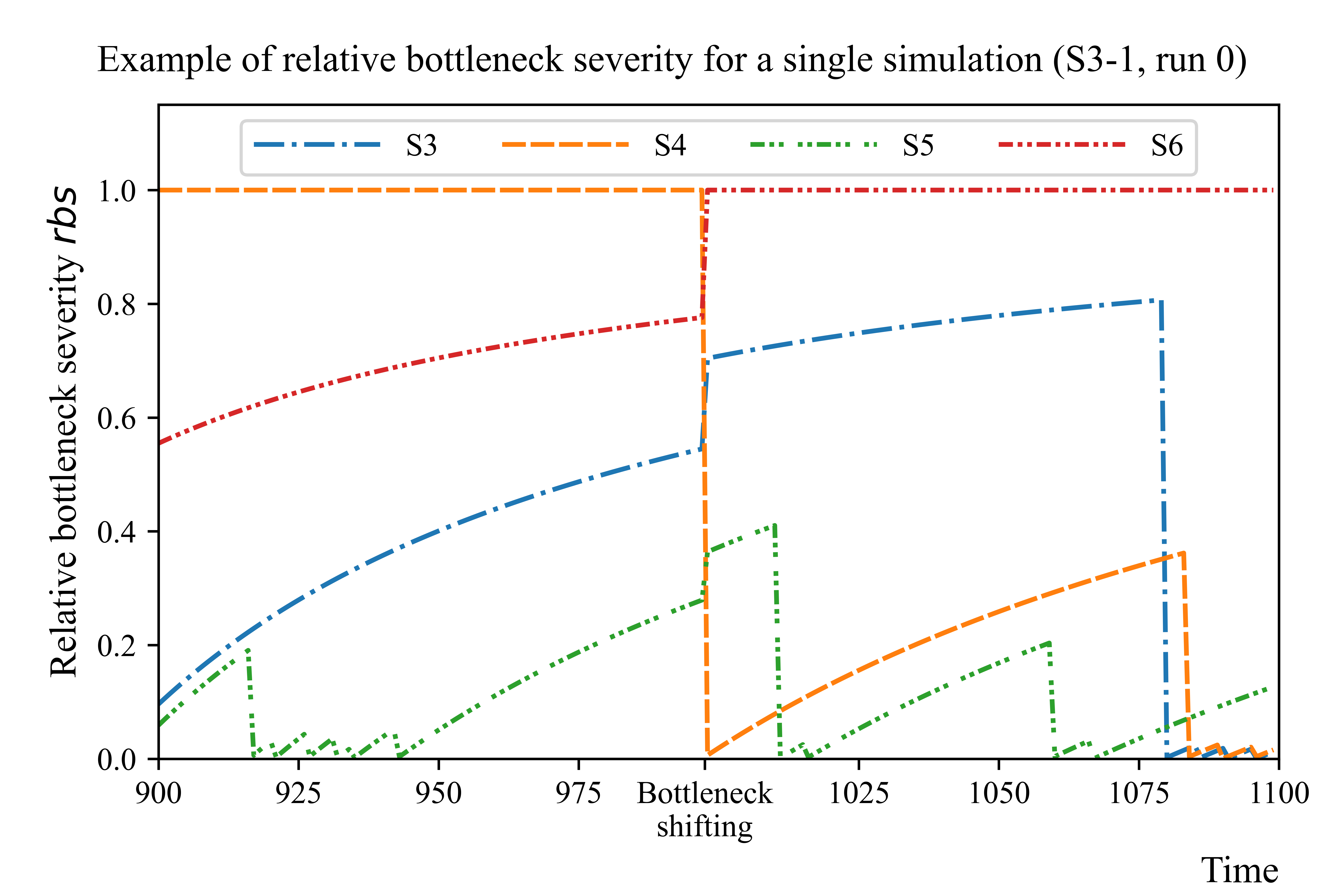}
  \caption{Exemplary view of the $rbs$ in $S3\mbox{-}1$ for a time interval between $900$ and $1100$}
  \label{fig:fig9}
\end{figure}

The benefit of the $rbs$ becomes apparent when viewed in individual time steps. \textbf{Figure \ref{fig:fig9}} shows a period of a bottleneck shift taking place. For the sake of clarity, we limit the visualization to stations $S3$ to $S6$. At a time of about 997, the current bottleneck shifts from $S4$ to $S6$. Thus, the $rbs$ of $S4$ drops to 0 and the $rbs$ of $S6$ changes to 1. The other stations change relative to the $rbs$ of $S6$. Consequently, in a consideration in momentary terms, $S4$ becomes the best target for immediate improvement measures up to time of 997, while $S6$ is already emerging as the successor bottleneck, marked by a comparatively high $rbs$ value.

\section{Conclusion}
\label{sec:conclusion}

With $rbf$ and $rbs$, two novel metrics for the diagnosis of dynamic bottlenecks were proposed in this paper. The metrics extend the field of bottleneck research, which has mainly been characterized by a focus on bottleneck detection and prediction. The metrics provide by a simple and practical way to quantify bottleneck behavior of a  manufacturing system. While relative bottleneck frequency can be used to evaluate long-term periods of observation, the relative bottleneck severity allows for an examination of current points in time. The code underlying the simulation and the subsequent evaluations for the calculation of the active periods, relative bottleneck frequency and relative bottleneck severity were made freely publicly available.

With regard to a promising possibility for the development of bottleneck metrics, we note at this point the not yet given possibility for the monetary evaluation of occurring bottlenecks. Only by determining availability losses in near-real time monitoring, it is possible to evaluate the monetary costs of bottleneck behavior. This provides an important argument when constructive or organizational improvement measures have to be selected and granted to deal with the bottlenecks. Thus, in a continuation of metrics for evaluating bottleneck states, a monetary quantification of throughput losses due to bottleneck events should be considered. This allows a consideration of the potential costs for the remedial measures of the bottlenecks and the costs due to the throughput losses that occur. 

Finally, we emphasize the untapped potential of bottleneck prescription. To implement intelligent system, which reacts independently to occurring bottlenecks with appropriate measures, a data-driven way for the evaluation of bottleneck conditions is required. Since such a system must also operate under resource constraints, the question of prioritizing emerging bottlenecks also arises here. In times of an increasing demand for sustainable and resource-efficient manufacturing systems, we foresee that the field of Bottleneck Analysis will continue to gain importance in the future.

\section*{Acknowledgments}
This paper is part of the project ‘Prediction of dynamic bottlenecks in directed material flow systems using machine learning methods’ (PrEPFlow, 21595), which is funded by the German Federal Ministry of Economics and Technology (BMWi), through the Working Group of Industrial Research Associations (AIF). It is carried out on behalf of the German Logistics Association e.V. (BVL) and it is part of the program for promotion of joint industrial research and development (IGF) based on a resolution of the German Bundestag.

\bibliographystyle{unsrt}  
\bibliography{references}  

\end{document}